\documentclass[journal]{ieeetran}

\usepackage{amssymb, amsmath, graphicx, paralist,subfigure,cite}
\usepackage{algorithm2e}
\usepackage{color}

\newtheorem{theorem}{Theorem}

\def\mb{\mathbf}
\def\mc{\mathcal}
\DeclareMathOperator*{\argmin}{argmin}

\begin{document}

\title{Cyber-Social Systems: Modeling, Inference, and Optimal Design}

\author{Mohammadreza Doostmohammadian$^\ast$,  Hamid R. Rabiee$^\dagger$, \textit{Senior Member, IEEE},  Usman A. Khan$^\ddagger$, \textit{Senior Member, IEEE}
	
\thanks{
	$^\ast$ Mechanical Engineering Department, Semnan University, Semnan, Iran, \texttt{doost@semnan.ac.ir}
	
	$^\dagger$ ICT Innovation Center for Advanced Information and Communication Technology, School of Computer Engineering, Sharif University of Technology, Tehran, Iran, {\ \texttt{rabiee@sharif.edu}}.
		
	$^\ddagger$ Electrical and Computer Engineering Department, Tufts University, Medford, USA \texttt{khan@ece.tufts.edu}.}}
\maketitle

\begin{abstract}
This paper models the cyber-social system as a cyber-network of agents monitoring states of individuals in a social network. The state of each individual is represented by a social node and the interactions among individuals are represented by a social link. In the cyber-network each node represents an agent and the links represent information sharing among agents. Agents make an observation of social states and perform \textit{distributed inference}. In this direction, the contribution of this work is threefold: (i) A novel distributed inference protocol is proposed that makes no assumption on the rank of the underlying social system. This is significant as most protocols in the literature only work on full-rank systems. (ii) A novel agent classification is developed, where it is shown that connectivity requirement on the cyber-network differs for each type. This is particularly important in finding the minimal number of observations and minimal connectivity of the cyber-network as the next contribution. (iii) The cost-optimal design of cyber-network constraint with distributed observability is addressed. This problem is subdivided into sensing cost optimization and networking cost optimization where both are claimed to be NP-hard. We solve both problems for certain types of social networks and find polynomial-order solutions.

\end{abstract}

\section{Introduction}\label{sec_introduction}

\begin{figure}[!hbpt]
	\centering
	\includegraphics[width=3in]{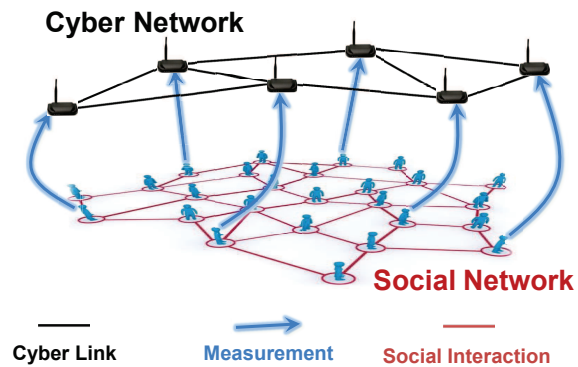}
	\caption{This figure shows a cyber-social system: a social system, represented by interaction of individuals over a social network, monitored by a cyber-network of agents. In this figure, the social network may represent a consensus network or dynamics of opinions. The blue links from the social nodes to the cyber nodes represent the measurement or observation taken from the social states of individuals by the agents, and the intelligent units tracking and observing the state of individuals connected via, for example, a wireless network represent the cyber-network. Based on observed information, agents share sufficient information and perform distributed inference, and therefore, are able to globally track the state of all individuals in the social network.} 
	\label{fig_cybersocial}
\end{figure}

\IEEEPARstart{C}{yber}-Social Systems (CSS) have recently gained considerable attentions in the literature \cite{rho2016cyber,khaitan2015design,wang2010emergence}. The typical structure of such systems, as a Cyber-Physical System (CPS), includes a cyber-network of agents  monitoring a physical system (that could be a social network
of individuals), see Fig.\ref{fig_cybersocial}.
A recently developed related concept is Cyber-Physical-Social Systems \cite{wang2010emergence} (CPSS)-- that is, the CPS tightly coordinated and integrated with social and human characteristics. In CPSS the coordination is three-fold: the cyber-network, the physical system, and the social or human network. However, in this work the social system, that may model the network of interactions among individuals, represents the physical system to be tracked by the cyber-network. In other words, we do not consider a different social system in parallel with the physical system.
The cyber-network represents the group of devices with  sensing, communicating, and data processing abilities embedded in every-day life of human individuals.
The social system typically models the interaction of individuals in a veriaty of contexts, e.g. collaboration, communities, economy, epidemic, etc.
What is uniform among all social examples is the structure of social network, or the \textit{social digraph} irrespective of dynamics or phenomena governing the network.

Note that the social graph structure dictates many properties of the overall cyber-social system. In this direction, we adopt a structural approach and model the social system as a Linear Structure-Invariant (LSI) model. In such models the system structure (zero/nonzero pattern) remains fixed while the non-zero values of the system matrix possibly change in time \cite{woude:03}. This approach is particularly beneficial in simplified models for nonlinear systems, as the Jacobian matrix of the nonlinear system is structure-invariant while the entries change based on the linearization point \cite{Liu_nature, nonlin}. It is known that, many properties of a LSI system are \textit{generic} (or structural) depending only on the system matrix structure, including observability \cite{liu_pnas} and controllability \cite{Liu_nature}. In such models the non-admissible choices for which the structural properties do not hold lie on a subspace with zero Lebesgue measure \cite{woude:03} and, therefore, it is typical to claim that such properties hold for \textit{almost-all } values of nonzero system parameters. This motivates the structural approach in this paper to explore the generic properties of the cyber-social network, namely state inference. In distributed inference, the states of individuals in the social network are tracked by agents, where agents communicate with each other to locally estimate the global state of the social network \cite{acc13,jstsp,jstsp14,commault_recovery,nuno-suff.ness,battistelli2011information,usman_cdc:10,Sayed-LMS}. Application of distributed inference in human social systems, for example, can be found in \cite{hao2009multiple}.
A potential application of this work might be deduced from \cite{wai2016active,wu2018estimating} based on the \textit{DeGroot model} for opinion dynamics \cite{degroot1974reaching}. Consider a group of agents or \textit{social sensors} taking the opinions of individuals interacting on a social network. Then the information is shared over a communication network among agents, and using an efficient distributed protocol enables the agents to track the global dynamic opinion of the community.
	
\textit{Contributions and related literature: }In this paper, we first study the distributed inference over the cyber-social network. In particular, we characterize \textit{distributed observability} as a requirement for distributed inference. The first challenge and motivating  question answered in this paper is: \textit{how to design the cyber-network structure and under what inference protocol one can achieve distributed inference of social digraph.} We are particularly concerned about the rank condition of the underlying social system. The approach in this paper only depends on the social interaction of individuals in the social digraph and  is independent of the particular social dynamics governing the system. In this direction, we characterize the social system as  LSI.

Note that our distributed inference protocol significantly differs from centralized  \cite{liu_pnas,doostmohammadian2017observational} and semi-centralized inference \cite{ sauter:09,commault_recovery}. In this paper, we provide a novel agent classification based on the type of state measured by the agent. This is further used to determine the minimum number of measurements for observability.
As compared to \cite{nuno-suff.ness,battistelli2011information,usman_cdc:10,Sayed-LMS} on distributed estimation, we  partition the agents (taking measurements) into Type-$\alpha$ and Type-$\beta$\footnote{We define these two types in Section~\ref{sec_obsrv}. These notions are closely related to the concepts of contraction and parent SCC. Type-$\alpha$ agent measures a state in a contraction, while Type-$\beta$ agent measures a  state in a parrent SCC.} and  show that distributed inference only needs Type-$\alpha$ measurements  directly at each agent, while Type-$\beta$ measurements are needed over a  path. This classification is significant as it provides the most general case based on matching properties of the social system and, further, it imposes less communication burden on the cyber-network as compared to, for example, \cite{commault_recovery}. We characterize distributed observability as a requirement for distributed inference. This notion is used later on the optimal design of cyber-network.

The next contribution  is to find the optimal solution for distributed observability. The motivation is to determine the sensing nodes and assign the agents such that the social inference cost is minimized, which has real-world applications \cite{dou2017optimizing}. In this direction, first the minimum observability requirements are addressed. Then the problem is subdivided into sensing cost optimization and networking cost optimization. In the sensing cost optimization the problem is the cost-optimal selection of  state measurements from social digraph. Note that any social state observation by agents is accompanied with a cost due to social conditions or even sensor embedding or installation cost. However, not all collection of state measurements gives an observable inference. With no observability the inference error goes unbounded. To solve the optimization, due to graph-theoretic nature of this problem, the structural observability constraint is considered.
Due to complexity  under observability constraint, the literature on this optimization problem is limited to \cite{pequito_gsip} (to the best of our knowledge), where the problem is claimed to be NP-hard\footnote{NP-hard problems are believed to have no solution in time-complexity upper-bounded by a polynomial function of the input parameters. Polynomial order algorithms are desirable in large-scale as their running time is upper-bounded by a polynomial expression in the size of algorithm input.} for general systems.

This paper finds a polynomial order solution of the optimization problem under observability constraint for \textit{matched} social digraphs. Note that the social digraph is matched if and only if it contains a disjoint union of family cycles spanning all nodes. An example is a \textit{self-damped social system} \cite{acc13_mesbahi,moothedath2018minimum} where the social graph includes self-edges at each node. This simply implies that individuals make their opinions based on previous states of the neighbors along with their own previous state. Example of such systems can be found, for example, in social epidemic models \cite{nowzari2016analysis}, and in ecosystem models \cite{may2001stability}.
The solution also holds if the sets of\textit{ contractions} and \textit{parent SCCs}\footnote{The sets of contractions and parent SCCs are later defined in Section~\ref{sec_obsrv}.} of the social digraph are disjoint. With these assumptions, we reformulate the problem as a Linear Sum Assignment Problem (LSAP) \cite{assignmentSurvey}. This is significant because the LSAP has a \textit{polynomial-order} solution.

Next, we consider the networking cost optimization.  The problem is  cost-optimal design of the  cyber-network to satisfy distributed observability. The motivation is to minimize the cost of communication among agents. This, for example, finds application in localization of wireless sensor networks embedded in real social networks. To the best of our knowledge, the cost-optimal design of cyber-network under distributed observability constraint is not considered in the literature, while the \textit{centralized} case is limited to \cite{pequito2017structurally}. Here, to solve the problem for distributed inference, we assume the underlying social digraph is matched\footnote{The general networking cost optimization problem is NP-hard \cite{doostmohammadian2018structural}.}. In this problem the cost of a cyber link may represent communication cost of agents. Among all possible links in cyber-network the problem is to select the cost-optimal links while satisfying distributed observability constraint. To solve the problem it is assumed that the communication links are \textit{bidirectional}. This is a logical assumption as if agent $i$ is in communication range of agent $j$ both agents communicate their information. With this assumption a combinatorial solution is provided for this problem.
This generalizes the centralized cost-optimal design problem in \cite{pequito2017structurally} to \textit{distributed} inference while the constraint is distributed observability.

In the followings we summarize the contribution of our work and the differences with the existing literature:
\begin{itemize}
\item Distributed inference papers \cite{nuno-suff.ness,battistelli2011information,usman_cdc:10,Sayed-LMS} (with partial observability of agents) in the literature assume that the underlying  system	is full-rank. This assumption significantly reduces the connectivity requirement in the cyber-network of agents. By this assumption, the papers \cite{nuno-suff.ness,battistelli2011information,usman_cdc:10,Sayed-LMS} simply assume that the cyber-network is  connected (if undirected) or strongly-connected (if directed). In this direction, our work significantly differs from the similar distributed inference protocols in the literature as it does not make any assumption on the rank of the underlying system and, therefore, it provides the general condition on minimal connectivity requirements for the general (full-rank or rank-deficient) systems.
\item Unlike some recent works in distributed inference \cite{kar2012distributed}, this work does not assume global observability of agents in their neighborhood. In \cite{kar2012distributed} it is assumed that for each agent the underlying system is globally observable by agent itself and its neighboring agents. This assumption  requires significantly more connectivity in the cyber-network as compared to our work. Similarly, semi-centralized inference protocols \cite{sauter:09,commault_recovery} require all-to-all cyber-network and  significant more connectivity.
\item The sensing cost optimization under structural observability constraint is claimed to be NP-hard in \cite{pequito_gsip}, while in Section~\ref{sec_sensingcost} we derive a polynomial order solution for (i) social systems with matched digraphs, and (ii) systems with disjoint set of parent SCCs and contractions. We further develop the optimal design of cyber-network under \textit{distributed} observability constraint and bidirectional links in the cyber-network. The optimal design of the network under distributed observability is not considered in literature. The work \cite{pequito2017structurally} is somewhat related, however they develop under \textit{centralized} estimation and observability and are not applicable for distributed inference.
\item As compared to our previous works \cite{doostmohammadian2017observational,doostmohammadian2018structural}, in this work we provide novel distributed inference protocol and agent classification for cyber-social systems and discuss the minimal sensing and cyber-network connectivity requirements. Further, we extend and generalize optimal design of cyber-network for matched social digraphs and further, social digraphs with disjoint sets of contractions and parent SCCs.
\end{itemize} 	

\textit{Assumptions:} In this paper, it is assumed that the structure of the social network is fixed, while the social parameters may change. The social system is assumed to be accompanied by Gaussian noise. The group of agents (all-together) have access to the global information of the social system, however, information of each agent alone is partial and local (not global).  The cyber links among agents are assumed to be noiseless and without packet loss, with no link failure. The reason is that we aim to find minimal sufficient network among agents. Perhaps, with consideration of noise and imperfections, the sufficient number of communication links in the cyber-network would increase.

\section{Social Dynamic Formulation}\label{sec_social}
\subsection{Social Model}
Social systems are modeled in both linear and nonlinear frameworks \cite{blondel2009krause ,friedkin2006structural}; examples are consensus (agreement) problems  \cite{Friedkin:1999vd,gossip_moura2010,pequito_gsip} and Markov-based opinion dynamics~\cite{degroot1974reaching,wu2018estimating,acemoglu2011opinion,castellano2009statistical}. However the  well-known models of social influence networks are proposed by Freidkin ~\cite{Friedkin:2003tx} and French~\cite{French}. This model is further developed in~\cite{Xing:2010ft}, where the social graph structure is fixed while the influence weights are time-varying. These references model the social system in state-space as LSI, where the social interaction of individuals are represented by a \textit{social digraph} with state nodes representing the attitudes, sentiments, opinions, or expectations of individuals, and graph links, with potentially time-varying weights, represent the social influence, trust, or faith of individuals on one another. It should be noted that the LSI model can be used as a simplified model of nonlinear systems where the structure of the Jacobian matrix is invariant while the entries are a function of linearization point \cite{nonlin}; consider the system dynamics to follow a nonlinear equation of states $f(\mb{x})$. The entries of the Jacobian matrix $\mc{J}$ are $\mc{J}_{ij}= \frac{\partial f_i}{\partial x_j}$. If $f_i$ is a function of $\mb{x}_j$, the entry $\mc{J}_{ij}$ is nonzero. In linearization model, the nonlinear function is linearized on the system operating point. The exact numerical value of $\mc{J}_{ij}$ depends on the operating point while the zero-nonzero structure of $\mc{J}$ is irrespective of the operating point. This is particularly the case in  observability or controllability analysis ~\cite{liu_pnas,Liu_nature}. Hence,  to model the social phenomena, it is natural to use the LSI model where the (nonzero) elements, representing the social influence weights, change in time as long as the system matrix structure, representing the social digraph, is not violated, e.g. see~\cite{Friedkin:2003tx} and~\cite{Xing:2010ft}.
Hence, many properties of social phenomena, given by nonlinear dynamics, can be analyzed on their LSI descriptions. We repeat that in this paper, we deal with the structure of the social networks, i.e. social interactions, as opposed to their specific dynamics. This structure may come from a linearization of nonlinear dynamics and results into an LSI system. Therefore, generic properties, including observability,  of the linearized version is structurally similar to the nonlinear system; see \cite{woude:03} for more information. Let us assume the state-vector of a social network to comprise of the opinions of certain individuals. If the opinion of person  "a" is dependent on the opinion of person  "b" via a very nonlinear equation, in the structured system representation, the entry at the row "a" and column "b" is nonzero and may change in time. This represents a generic (structural) setting, where the properties are irrespective of specific numerical analysis and only depends on the structure \cite{woude:03}. It should be noted, the LSI model outperforms the LTI system models as the LSI results are generic. Note that, for the LTI model the system properties, e.g. observability, must be checked \textit{numerically} for every choice of system entries. However, for LSI models the properties such as observability can be checked via efficient graph-theoretic algorithms and the results hold for almost all values of system entries while the structure is fixed. Therefore, the structural analysis outperforms the LTI (or other numerical methods) in terms of computational complexity. For typical social models, where the system entries are time-variant representing the change in social influence weights, the LSI model is better suitable for inference over cyber-social systems.
In this direction, model the social dynamics as,
\begin{eqnarray}\label{eq_sys1}
\mb{x}_{k+1} = A_k\mb{x}_k + \mb{v}_k,\qquad k\geq0,
\end{eqnarray}
where~$A_k$ is the system matrix,~$\mb{v}_k$ is Gaussian noise on states, and $\mb{x}_k\in\mathbb{R}^n$ is the social state vector defined as,
\begin{eqnarray}\nonumber
\mb{x}_{k} &=& \left( \begin{array}{c}
\mb{x}_{k}^1\\
\vdots \\
\mb{x}_{k}^n
\end{array}\right)
\end{eqnarray}
$\mb{x}_{k}^i$ is the state of the $i$th individual in the social network. $n$ is the number of social states, i.e. the size of the social network.
Further, $\mc{A} \in  \{0,1\}^{n \times n}$ denotes the $0-1$ structure of system matrix $A_k$; in other words every nonzero entry of $A_k$ is represented by $1$ in $\mc{A}$ and fixed zeros of $A_k$ are represented by $0$ in $\mc{A}$. The $\mc{A}$ structured matrix may be inferred from, for example, voting records using discrete random processes as described in \cite{wu2018estimating}.

Consider a cyber-network of agents monitoring the social dynamics \eqref{eq_sys1} by taking measurements as:
\begin{eqnarray} \label{eq_H_i}
\mb{y}_k^i = H_i\mb{x}_k + \mb{r}_k^i.
\end{eqnarray}
where~$H_i\in\mathbb{R}^{p_i\times n}$ is the local measurement matrix at agent~$i$ and time~$k$; $\mb{y}^i_k\in\mathbb{R}^{p_i}$  is the vector representing local measurement (where $p_i$ is the number of measurements by agent $i$), and~$\mb{r}^i_k$ represents noise (with standard assumption on Gaussianity\footnote{Gaussian noise is a standard assumption in the literature towards optimal state estimation. When the noise is not Gaussian, the optimal minimum mean squared estimator may not have a simple recursive representation as in Kalman filtering. In our case, the Gaussian assumption is not needed as the observability analysis is impervious to noise. We have provided noise statistics for the sake of completeness. Knowing that one may potentially lose the optimality of the ensuing estimator, the Gaussian assumption can be removed and it is typical to replace this by noise with zero-mean and a bounded variance, see \cite{kar2009distributed} for example.}). Then, the global measurement model is,
\begin{eqnarray}\nonumber
\left(
\begin{array}{c}
\mb{y}^1_{k}\\
\vdots\\
\mb{y}^N_{k}
\end{array}
\right) &=&
\left(
\begin{array}{c}
H_{1}\\
\vdots\\
H_{N}
\end{array}
\right)\left(
\begin{array}{c}
\mb{x}^1_{k}\\
\vdots\\
\mb{x}^n_{k}
\end{array}
\right)+
\left(
\begin{array}{c}
\mb{r}^1_{k}\\
\vdots\\
\mb{r}^N_{k}
\end{array}
\right),\\
\triangleq \mb{y}_k &=& H\mb{x}_k + \mb{r}_k,
\end{eqnarray}
where~$\mb{y}_k\in\mathbb{R}^{p}$ is the global measurement vector,~$H=\{H_{ij}\}\in\mathbb{R}^{p\times n}$ is the global measurement matrix, $N$ is number of agents, and~$\mb{r}_k$ represents noise (with Gaussian assumption). Note that~$p=p_1+\ldots+p_N$.

Agents,~$i=1,\ldots,N$, monitoring the social network, exchange information over the cyber-network represented by graph,~$\mc{G}_W=(\mc{V}_W,\mc{E}_W)$, where,~$\mc{V}_W$, represent the set of all agents, and,~$\mc{E}_W$, is the set of links. Each link is represented by~$(i,j)$, implying that agent~$j$ can send information to agent~$i$ over the communication channel. $\mc{N}_i$ describes the neighborhood of agent~$i\in\mc{V}_W$  as,
\begin{eqnarray}
\mc{N}(i)=\{j~|~(i,j)\in\mc{E}_W\},
\end{eqnarray}

\subsection{Problem Formulation}

\textit{As one of the contributions of this paper, we are particularly focused on designing the structure of $H$ matrix, i.e. which states to be observed by the agents. Further, we are interested in the structure of adjacency matrix of $\mc{G}_W$, i.e. how the agents link over the cyber-network, given only the pattern of $A$ as the structure of social digraph. As the next problem, the structure of cyber-network $\mc{G}_W$ needs to be designed such that: (i) the tracking error at each agent is bounded steady-state, i.e. the social system is observable by each agent; and (ii) the cost of sensing the social state by each agent and cost of communicating inference information with other agents over the cyber-network is minimized.}

To solve the problem, we use structured systems theory. In this setting, Eqs.~\eqref{eq_sys1}-\eqref{eq_H_i} are modeled by a \emph{system digraph}, where the graph nodes $\mc{X}\triangleq\{x_1,\ldots,x_n\}$ are states and the graph edges represent the (social) interactions by structure of~$A$ (or structured matrix $\mc{A}$) as ~$\mc{E}_A=\{(x_i,x_j)~|~a_{ij}\neq0\}$ (interpreted as~$x_i\leftarrow x_j$) and $\mc{G} = (\mc{X},\mc{E}_A)$.  Define~$\mc{Y}\triangleq\{\mb{y}^1,\ldots,\mb{y}^N\}$ denoting the set of measurements by agents\footnote{In this paper without loss of generality it is assumed that each agent takes one measurement of social system.}.
Further define~$\mc{G}_{\scriptsize \mbox{sys}} = (\mc{V}_{\scriptsize \mbox{sys}},\mc{E}_{\scriptsize \mbox{sys}})$, described on both states and measurements:~$\mc{V}_{\scriptsize \mbox{sys}}=\{\mc{X} \cup \mc{Y}\}$, and~$\mc{E}_{\scriptsize \mbox{sys}}=\{(x_i,x_j)~|~a_{ij}\neq0\} \cup \{(y_i,x_j)~|~h_{ij}\neq0\}$. Indeed~$\mc{G}_{\scriptsize \mbox{sys}}$, represents the pair~$(A,H)$, where the nodes consists of both individual state nodes and measurements (agents).

\section{Structural Observability and Related Graph  Notions} \label{sec_obsrv}
In this section we discuss the observability of the social digraph by the cyber-network of agents. For this purposes, we first define a \emph{$\mc{Y}$-connected} path,~$i\overset{\scriptsize\mbox{path}}{\longrightarrow} j$, from~$i\in\mc{X}$ to~$j\in\mc{Y}$, as a sequence of state nodes,~$\{i,{i_1},\ldots,i_{l-1},j\}$, with~$(j,i_{l-1}),\ldots,(i_1,i)\in\mc{E}_{\scriptsize \mbox{sys}}$. With this notation we state the main theorem on structural observability over graphs:
\begin{theorem}\label{th1}
	A system is structurally observable if and only if in its digraph,~$\mc{G}_{\scriptsize \mbox{sys}}$:
	\begin{inparaenum}[(i)]
		
		\item (\textbf{Accessibility})  for every~$i\in\mc{X}$ there is a $j\in\mc{Y}$ such that $i\overset{\scriptsize\mbox{path}}{\longrightarrow} j$, i.e. all state nodes are $\mathcal{Y}$-connected.
		
		\item (\textbf{Rank condition}) There exist a family of \textit{disjoint}~cycles and~$\mathcal{Y}$-connected paths spanning all  state nodes $\mc{X}$.
	\end{inparaenum}
\end{theorem}
The proof is provided in \cite{liu_pnas}.\footnote{Note that based on \cite{woude:03}, such observability holds for almost all numerical values of system parameters.}
In light of Theorem~\ref{th1}, the concepts of Contractions and Strongly Connected Components (SCC) are tied with structural observability.  In the coming sections we show that observation of (at least) one state node in each of these components are required for observability and inference.

\subsection{Contractions}\label{sec_cont}
Contractions are better defined over the bipartite representation of the digraph $\mc{G}=(\mc{X},\mc{E}_A)$. Bipartite graph,~$\Gamma=(\mc{V}^+,\mc{V}^-,\mc{E}_\Gamma)$, is constructed by two disjoint set of nodes:~$\mc{V}^+=\mc{X}$ and~$\mc{V}^-=\mc{X}$, with all edges~$\mc{E}_\Gamma$ from~$\mc{V}^+$ to~$\mc{V}^-$. The  bipartite graph,~$\Gamma$, is constructed from ~$\mc{G}$ with the edge set~$\mc{E}_{\Gamma}$, defined as the collection of~$(v_j^-,v_i^+)$, if ~$(v_j,v_i) \in \mc{E}_A$. Define matching,~$\underline{\mc{M}}$, on the social graph,~$\mc{G}$, as the subset of edges,~$\mc{E}$, having no common end node, which defines the subset of edges no two of them incident on the same node in $\mc{V}^+$ in bipartite graph,~$\Gamma$. This implies a set of disjoint edges in~$\underline{\mc{M}}$. Define the matching size,~$|\underline{\mc{M}}|$, as the number of edges in $\underline{\mc{M}}$, and further, a matching having maximum possible size as maximum matching, denoted by~$\mc{M}$. Denote the nodes incident to~$\mc{M}$ in~$\mc{V}^+$ and~$\mc{V}^-$ by $\partial \mc{M}^+$ and~$\partial \mc{M}^-$ respectively. Further, define~$\delta \mc{M} = \mc{V}^+ \backslash \partial \mc{M}^+$. Next, auxiliary graph,~$\Gamma^\mc{M}$, is constructed from bipartite graph,~$\Gamma$ and maximum matching,~$\mc{M}$,  by keeping the direction of all edges~$\mc{E}_{\Gamma} \backslash \mc{M}$ while  reversing the direction of edges in~$\mc{M}$. In the auxiliary graph,   define ${\mc{M}}$-alternating path as a sequence of edges starting from an unmatched node in~$\delta \mc{M}$ and every second edge in~$\mc{M}$. Denote ${\mc{M}}$-alternating path by $\mc{Q}_{\mc{M}}$. In the auxiliary graph, ${\mc{M}}$-augmenting path, $\mc{P}_{\mc{M}}$, is an ${\mc{M}}$-alternating path with begin and end node in $\delta \mc{M}$. Finally, having defined these notions we are ready to introduce the concept of contraction. \textit{In the auxiliary graph representation,~$\Gamma^\mc{M} _A$, for every unmatched node~$v_j \in \delta \mc{M}$ define a contraction set, denoted by $\mc{C}_i$, as  all state nodes in $\mc{V}^+$ reachable by ${\mc{M}}$-alternating paths starting from~$v_j$}.
The above constructions to find contractions are summarized in algorithm~\ref{alg_cont}.
\begin{algorithm} \label{alg_cont}
	\textbf{Given:} Social digraph $\mc{G}$
	
	Construct $\Gamma$\;
	Find a matching $\underline{\mc{M}}$ \;
	Construct auxiliary graph $\Gamma^{\underline{\mc{M}}}$ \;
	\While{augmenting path $\mc{P}_{\underline{\mc{M}}}$ exist}{
		\For{nodes in $\delta \underline{\mc{M}}$}{
			Find $\mc{P}_{\underline{\mc{M}}}$ \;
			$\underline{\mc{M}} = \underline{\mc{M}} \oplus \mc{P}_{\underline{\mc{M}}}$ \;
		}
	}
	Construct auxiliary graph $\Gamma^{\mc{M}}_A$ \;
	\For{nodes in $\delta \mc{M}$}{
		Find alternating paths $\mc{Q}_{\mc{M}}$ in $\Gamma^{\mc{M}}_A$ \;
		Put all nodes in $\mc{V}^+$ reachable by $\mc{Q}_{\mc{M}}$ in $\mc{C}_i$\;}
	
	\textbf{Return} $\mc{C}_i, i = \{1,...,l\}$\;\
	
	\caption{This algorithm finds contractions in social digraph.}
\end{algorithm}

Note that contractions play a key role in estimation recovery \cite{doostmohammadian2017distributed,doostmohammadian2017observational}.
\textit{It can be proved that measurement of the nodes in the same contraction improves the observability rank-condition by one \cite{icassp16}.} In fact, it is known that S-rank of matrix~$\left[ \begin{array}{cc} A^\top & H^\top \end{array} \right]^\top$ equals the size of maximum matching~$\mc{M}$ in digraph~$\mc{G}_{sys}$ and adding a measurement of an unmatched node in ~$\delta \mc{M}$ recovers this rank by one \cite{Liu_nature,jstsp}. Further notice that all contractions are distinct sets and do not share nodes. The nodes in the same contraction are indeed \textit{equivalent} unmatched nodes for different choices of maximum matching~$\mc{M}$. Therefore, the increase in the S-rank by observing nodes in the same contraction is one. In other words, if $H_{\mc{C}_i}$ represent measurement matrix of state nodes in a contraction~$\mc{C}_i$, then
\begin{eqnarray}
S\mbox{-rank}\left(
\begin{array}{c} A \\ H_{\mc{C}_i}
\end{array} \right) = S\mbox{-rank}~(\mc{A}) +1.
\end{eqnarray}
We call measurement of a state node in a contraction as \textit{Type-$\alpha$ measurement}. Also the agent measuring such state node is called Type-$\alpha$.
\subsection{SCCs}
In a social digraph, a Strongly Connected Component (SCC), represented by~$\mc{S}$, is defined as maximal strongly-connected partitions in $\mc{G}=(\mc{X},\mc{E}_A)$. Note that strong connectivity means that there exist a directed path from every node to every other node in that partition, i.e. $x_i\overset{\scriptsize\mbox{path}}{\longrightarrow} x_j$ for every $x_i,x_j$ in~$\mc{S}$. An SCC is called \textit{parent}, denoted by~$\mc{S}^{p}$, if it has no outgoing edges to any other state in $\mc{G}\setminus \mc{S}^{p}$. Any non-parent SCC is called \textit{child} SCC, denoted by~$\mc{S}^{c}$. Note that the SCC classification of a digraph and its parent or child partial order is performed via well-known algorithms, to name a few \textit{Tarjan’s
algorithm}~\cite{tarjan} and \textit{Depth-First-Search (DFS) algorithm}~\cite{algorithm}. We do not repeat the algorithms here and refer the interested readers to the mentioned references.

\textit{It can be proved that measurement of the nodes in the same parent SCC  recovers the accessibility condition   of structural observability \cite{icassp16}.} This is because all states in the same SCC are reachable  to each other via a path, and therefore, measurement of any state implies accessibility of all other states in that SCC to the measurement. In other words, for any $\{x_i,x_j\} \in \mc{S}^p$, having $x_i \overset{\scriptsize\mbox{path}}{\longrightarrow} \mc{Y}$ implies $x_j \overset{\scriptsize\mbox{path}}{\longrightarrow} \mc{Y}$ because $x_j \overset{\scriptsize\mbox{path}}{\longrightarrow} x_i$.
Note that child SCCs are not needed for accessibility. The reason is that for every states in a child SCC $\mc{S}^c_i$ there are paths to states in a parent SCC $\mc{S}^p_j$, and therefore $\mc{S}^c_i \overset{\scriptsize\mbox{path}}{\longrightarrow}  \mc{S}^p_j \overset{\scriptsize\mbox{path}}{\longrightarrow} \mc{Y} $.
For convenience, we name measurement of states in parent SCC as \textit{Type-$\beta$ measurement}. Also, the agent observing such state is called Type-$\beta$.

\section{Distributed Inference via Cyber Network of Agents}\label{sec_distributed}
Based on the notions in the previous section the main theorem on structural observability is stated here:

\begin{theorem} \label{thm_centralized}
	The necessary measurements for structural observability of a digrpah,~$\mc{G}$,  are:
	\begin{enumerate} [(i)]
		\item one state node from every contraction,~$\mc{C}$;
		\item one state node from every parent SCC,~$\mc{S}^{p}$.
	\end{enumerate}
\end{theorem}	
The proof of this theorem directly follows the statements in the previous section. We refer interested readers to \cite{jstsp,Liu_nature}.
Note that Theorem~\ref{thm_centralized} not only holds for centralized inference, but also the distributed case. Consider a cyber-network of agents monitoring the state of the social digraph. The global information of all social states can be estimated at each agent by having sufficient observations for observability. This can be done by directly sharing all necessary measurements via the cyber-network such that each agent has all sufficient measurements in its neighborhood \cite{commault_recovery,kar2012distributed,sauter:09}.  However, this method requires a large amount of communication among agents and a dense cyber-network. In general distributed cases, agents have access to partial measurements (even in their neighborhood) and thus, the idea is by sharing information (measurements or predictions) among the agents one can satisfy these  conditions  to recover  partial observability.
\textit{In this section, assuming no information loss over the links, the main objective is to first design the distributed inference protocol and then the conditions on the cyber-network of agents to ensure bounded estimation error, which is known as distributed observability}. Note that the assumption on no information loss among agents is a typical assumption in the sensor network and distributed estimation literature. Based on the recent technological advances in communication networks (used, for example, in wireless networks, cloud computing, and internet of things) this is a fair assumption.

The proposed distributed inference protocol in this paper consists of two steps of information sharing: (i) prediction sharing over graph $\mc{G}_\beta$, and (ii) measurement sharing over graph $\mc{G}_\alpha$; the combination of these two graphs make the cyber-network, i.e. $\mc{G}_W = \mc{G}_\beta \cup \mc{G}_\alpha $. Further define the neighborhood of agents (including each agent itself) $\mathcal{N}_\beta$ and $\mathcal{N}_\alpha$ respectively in $\mc{G}_\beta$ and $\mc{G}_\alpha $. We first consider prediction sharing over network $\mc{G}_\beta$ as follows:
\begin{eqnarray}\label{eq_p}
\widehat{\mb{x}}^i_{k|k-1} = \sum_{j\in\mathcal{N}_\beta(i)} w_{ij}A_k\widehat{\mb{x}}^j_{k-1|k-1},
\end{eqnarray}
where $W = \{w_{ij}\}$ denotes the adjacency  matrix of $\mc{G}_\beta $ and is a stochastic matrix. $\widehat{\mb{x}}^i_{k|k-1}$ represents the estimate of state,~$\mb{x}$, at time $k$, using all state predictions at agent~$i$, and its neighboring agents $\mathcal{N}_\beta(i)$ at time~$k-1$. $\widehat{\mb{x}}^j_{k-1|k-1}$ is the state estimate of neighboring agent $j$ at time $k-1$ given all its neighboring measurements at time $k-1$.  At the  next step we consider sharing the measurements among agents over the network $\mc{G}_\alpha$ as follows:
\begin{eqnarray}\label{eq_m}
\widehat{\mb{x}}^i_{k|k} =\widehat{\mb{x}}^i_{k|k-1} + K_k^i \sum_{j\in \mc{N}_\alpha(i)}H_j^\top \left(\mb{y}^j_k-H_j\widehat{\mb{x}}^i_{k|k-1}\right).
\end{eqnarray}
where $\widehat{\mb{x}}^i_{k|k}$ is the state estimate of agent $i$ at time-step $k$ given all the measurements of its neighboring agents $\mc{N}_\alpha(i)$ at time $k$. The gain matrix,~${K}_k$, is block-diagonal, i.e.~${K}_k=\mbox{blockdiag}[K_k^i,\ldots,K_k^N]$ and $K_k^i$ is the gain matrix at agent $i$. \textit{Note that block-diagonal ${K}_k$ is necessary to  ensure that the inference protocol is distributed.} However,  ${K}_k$, cannot be computed locally via the standard procedures in Kalman-type estimation. This constrained  gain matrix (to be block-diagonal) may be computed via an iterative cone-complementarity optimization algorithm based on Linear Matrix Inequality (LMI) approach \cite{rami:97}.

It should be noted, the inference protocol \eqref{eq_p}-\eqref{eq_m} do not impose any condition on the rank of $A$, as compared to the works \cite{nuno-suff.ness,battistelli2011information,usman_cdc:10,Sayed-LMS} which require $A$ to be full-rank. Further, in \eqref{eq_p}-\eqref{eq_m} for each agent $i$ there is no restricting  condition on the global observability of measurements $\mb{y}^j$ in the neighborhood $j\in \mc{N}_\alpha(i)$; this contradicts the work \cite{sauter:09,commault_recovery,kar2012distributed} which require global observability in the neighborhood of each agent.
	
Next, define the distributed estimation error as,
\begin{eqnarray}\label{love}
\mb{e}_{k}^i &=& \mb{x}_{k|k} - \widehat{\mb{x}}^i_{k|k}, \nonumber
\\ \nonumber
\mb{e}_{k}^i &=&\mb{x}_{k} - (\widehat{\mb{x}}^i_{k|k-1} + K_k^i \sum_{j\in \mc{N}_\alpha (i)}H_j^T (y^j_k-H_j\widehat{\mb{x}}^i_{k|k-1}))
\end{eqnarray}
Using equations \eqref{eq_p}-\eqref{eq_m} we have,
\begin{eqnarray}
\mb{e}_{k}^i &=&\mb{x}_{k} - (\sum_{j\in\mathcal{N}_\beta(i)} w_{ij}A\widehat{\mb{x}}^j_{k-1|k-1} \nonumber \\ &+& K_k^i \sum_{j\in \mc{N}_\alpha (i)}H_j^T
 (y^j_k-H_j\sum_{j\in\mc{N}_\beta (i)} w_{ij}A\widehat{\mb{x}}^j_{k-1|k-1})) \nonumber
\end{eqnarray}
Replacing the system equations \eqref{eq_sys1}-\eqref{eq_H_i} we get,
\begin{eqnarray}\nonumber
\mb{e}_{k}^i &=& (A\mb{x}_{k-1}+\mb{v}_{k-1}) - (\sum_{j\in\mathcal{N}_\beta(i)} w_{ij}A\widehat{\mb{x}}^j_{k-1|k-1}  \nonumber \\
&+& K_k^i \sum_{j\in \mathcal{N}_\alpha(i)}H_j^T (H_j\mb{x}_{k}+ \mb{r}^i_{k} \nonumber \\
&-& H_j\sum_{j\in\mathcal{N}_\beta(i)} w_{ij}A\widehat{\mb{x}}^j_{k-1|k-1})) \nonumber \\
\mb{e}_{k}^i &=&(A\mb{x}_{k-1}+\mb{v}_{k-1}) - \sum_{j\in\mathcal{N}_\beta(i)} w_{ij}A\widehat{\mb{x}}^j_{k-1|k-1} \nonumber \\
&-&K_k^i \sum_{j\in \mathcal{N}_\alpha(i)}H_j^T (H_j(Ax_{k-1}+\mb{v}_{k-1}) + \mb{r}^i_{k} \nonumber \\
&-& H_j\sum_{j\in \mathcal{N}_\beta(i)} w_{ij}A\widehat{\mb{x}}^j_{k-1|k-1}) \nonumber \\
\mb{e}_{k}^i &=&A\mb{x}_{k-1} - \sum_{j\in\mathcal{N}_\beta(i)} w_{ij}A\widehat{\mb{x}}^j_{k-1|k-1} \nonumber \\
&-& K_k^i \sum_{j\in \mathcal{N}_\alpha(i)}H_j^T(H_jAx_{k-1} - H_j\sum_{j\in \mathcal{N}_\beta(i)} w_{ij}A\widehat{\mb{x}}^j_{k-1|k-1}) \nonumber \\
&+& \mb{v}_{k-1}-\sum_{j\in \mathcal{N}_\alpha(i)}H_j^TH_j\mb{v}_{k-1} - \sum_{j\in \mathcal{N}_\alpha(i)}H_j^T\mb{r}^i_{k} \nonumber
\end{eqnarray}
Collecting the noise terms in a new parameter~$\mb{q}_k$ we have,
\begin{eqnarray}
\mb{e}_{k}^i &=& A\mb{x}_{k-1}-\sum_{j\in\mathcal{D}_i}\widehat{\mb{x}}^j_{k-1|k-1} \nonumber \\
&-& K_k^i \sum_{j\in \mathcal{N}_\alpha(i)}H_j^TH_j(Ax_{k-1} - \sum_{j\in\mathcal{N}_\beta(i)} w_{ij}A\widehat{\mb{x}}^j_{k-1|k-1}) \nonumber \\
&+& \mb{q}_k
\end{eqnarray}
Recall that matrix~$W$ is stochastic then we get,
\begin{eqnarray}
A\mb{x}_{k-1}=\sum_{j\in \mathcal{N}_\beta(i)} w_{ij}A\widehat{\mb{x}}^j_{k-1|k-1} \nonumber
\end{eqnarray}
and thus,
\begin{eqnarray}
\mb{e}_{k}^i &=& \sum_{j\in \mathcal{N}_\beta(i)}w_{ij}A(\mb{x}_{k-1} - \widehat{\mb{x}}^j_{k-1|k-1})  \nonumber \\
&-& K_k^i \sum_{j\in \mathcal{N}_\alpha(i)}H_j^TH_j\sum_{j\in \mathcal{N}_\beta(i)}w_{ij}A(\mb{x}_{k-1}-\widehat{\mb{x}}^j_{k-1|k-1}) \nonumber \\ &+& \mb{q}_k
\end{eqnarray}
Define the vector error $\mb{e}_{k}$ at all agents as 
\begin{eqnarray}\nonumber
\mb{e}_{k} &=& \left( \begin{array}{c}
\mb{e}_{k}^1\\
\vdots \\
\mb{e}_{k}^N
\end{array}\right)
\end{eqnarray}

Finally, the distributed inference error dynamics is as follows,
\begin{eqnarray}\label{eq_err1}
\mb{e}_{k} = (W\otimes A - K_kD_H(W\otimes A))\mb{e}_{k-1} +
\mb{q}_k,
\end{eqnarray}
where $D_H$ is defined as,
\begin{eqnarray}
D_H \triangleq \left(
\begin{array}{cccc}
\sum_{j\in \mathcal{N}_\alpha(i)}H_1^TH_1\\
&\ddots\\
& &\sum_{j\in \mathcal{N}_\alpha(i)}H_N^TH_N\
\end{array}
\right)
\end{eqnarray}
Based on Kalman filtering theory \cite{kalman:61} this equation is steady-state stable if and only if the pair $(W \otimes A, D_H)$ is observable, referred as \textit{distributed observability}.

At this point in the paper we develop conditions on the cyber-network~$\mc{G}_W$ (more precisely networks $\mc{G}_\beta$ and $\mc{G}_\alpha$) such that to satisfy distributed $(W\otimes A, D_H)$ observability. The results are based on graph notions  (contractions and SCCs), and also Type-$\alpha/\beta$ classification of agents. Note that for recovering distributed observability agents can  either share measurement $y_k^i$ or prediction $\widehat{\mb{x}}^i_{k|k-1}$. In this regard, we state the main theorem on (structural) distributed observability.

\begin{theorem}\label{thm_dist}
	The sufficient conditions for each agent,~$i$, to recover observability of   social graph~$\mc{G}$ is:
	\begin{enumerate}[(i)]
		\item For every contraction,~$\mc{C}_l$, agent~$i$ receives measurement information via a direct link from a Type-$\alpha$ agent measuring a state in~$\mc{C}_l$;
		\item For every parent SCC,~$\mc{S}^{p}_l$, agent~$i$ shares prediction information via a \textit{path} of links from a Type-$\beta$ agent measuring a state in~$\mc{S}^{p}_l$;
	\end{enumerate}
\end{theorem}

We provide a graph-theoretic approach for the proof. Note that part (i) recovers S-rank condition at every agent $i$. Every agent $i$ directly receives measurement of (a state in) all contractions measured by agents. Recall that having measurement of contraction $\mc{C}_l$ recovers the S-rank by one,
\begin{eqnarray}
S\mbox{-rank}\left(
\begin{array}{c} A \\ H_{\mc{C}_l}
\end{array} \right) = S\mbox{-rank}~(\mc{A}) +1.
\end{eqnarray}
Therefore, by having measurements from all contractions the S-rank deficiency of social matrix $A$ is recovered, satisfying the rank condition in Theorem~\ref{thm_centralized}. Part (ii) recovers the accessibility of parent SCC, $\mc{S}^{p}_l$. Note that  every agent $j$ measuring a state in a parent SCC $\mc{S}^{p}_l$ shares its prediction with agent $i$ via a path. Therefore, the parent SCC $\mc{S}^{p}_l$ is connected to output of agent $i$, i.e. $\mc{S}^{p}_l \overset{\scriptsize\mbox{path}}{\longrightarrow} y_i$ and therefore is $\mc{Y}$-connected. This holds for every agent $i$ and every parent SCC $\mc{S}^{p}_l$. Thus, the $\mc{Y}$-connectivity of every parent SCC is satisfied for every agent $i$, satisfying accessibility condition in Theorem~\ref{thm_centralized}. Having both conditions satisfied,  observability is satisfied at every agent.

Notice that condition (i) in Theorem~\ref{thm_dist} defines network~$\mc{G}_\alpha$, for measurement sharing, whereas condition (ii) defines network,~$\mc{G}_\beta$, over which agents \textit{only} share their state predictions.
Based on condition (i) $\mc{G}_\alpha$ is a network of hubs with $\alpha$-agents as hubs, and~$\mc{G}_\beta$ is a Strongly Connected (SC) network.
Note that the connectivity requirement in condition (ii) is weaker than the condition in~\cite{nuno-suff.ness}, where each agent requires to share  \textit{both} measurements and state predictions over the same network. Further, the conditions in Theorem~\ref{thm_dist} is weaker than network connectivity in~\cite{sauter:09}, which requires long-distance direct links for Type-$\beta$ agents. Note that in case the network $\mc{G}$ is matched, there is no contraction in the network and SC network $\mc{G}_\beta$ is sufficient for observability~\cite{usman_cdc:11, battistelli2011information}.

\section{Cost Optimal Design}\label{sec_optimal}

\subsection{Minimum Observability Requirement}\label{sec_minobsrv}
Recall that from Theorem~\ref{thm_centralized} we require (at least) one agent measuring a state in a contraction and one agent measuring a state in a parent SCC. It should be noted, however, contractions may share node with parent SCCs. Note that measurement of shared nodes of contractions and parent SCCs recovers both S-rank and accessibility conditions in Theorem~\ref{th1}. Therefore, for (distributed) observability, the minimum number of measurements (and in turn agents) is equal to:
\begin{equation}
N_{min}=|\mc{C}| +|\mc{S}^{p}|-|\mc{S}^{p} \cap \mc{C}|
\end{equation}
The measurement of the shared state between a parent SCC and a contraction is Type-$\alpha$. This is because sharing such measurement recovers both conditions (S-rank and accessibility) of the social digraph observability. However, as a Type-$\beta$, sharing the prediction of agent only recovers accessibility condition.

In case the social digraph $\mc{G}$ is structurally full-rank, there is no contraction in the social digraph. Therefore, for (distributed) observability, the minimum number of agents  is:
\begin{equation}
N_{min}=|\mc{S}^{p}|
\end{equation}
\subsection{Sensing Cost Optimization} \label{sec_sensingcost}
Assume that the observation by agents is associated with a cost. In sensing cost optimization the problem is to define the structure of $H$ such that the global cost is minimized. In other words, the cost-optimal selection of agents under observability constraint is addressed. This problem for general social graphs is claimed to be NP-hard in \cite{pequito_gsip}.\textit{ As another contribution, here we show that if the social graph is matched or it contains disjoint set of contractions and parent SCCs  this problem has a polynomial order solution}. Mathematically, the problem is,
\begin{equation} \label{eq_formulation1}
	\begin{aligned}
	\displaystyle
	\argmin
	\limits_{\mc{H}} ~~ & \sum_{i=1}^{N} \sum_{j=1}^{n} c_{ij}\mc{H}_{ij} \\
	\text{s.t.} ~~ & (A,H)\mbox{-structural~observability},\\
	~~ &  \mc{H}_{ij} \in \{0,1\}\\
	\end{aligned}
\end{equation}
where $c_{ij}$ is the cost of measuring state $j$ by agent $i$. Here, we describe the solution for matched social digraphs, and the solution for social digraphs with disjoint set of contractions and SCCs similarly follows. Note that the minimum number of agents for matched digraphs equals the number of parent SCCs and one state measurement from every parent SCC is sufficient for social inference. This results the following formulation:

\begin{equation} \label{eq_formulation3}
	\begin{aligned}
	\displaystyle
	\argmin
	\limits_{\mc{H}} ~~ & \sum_{i=1}^{N} \sum_{j=1}^{n} c_{ij}\mc{H}_{ij} \\
	\text{s.t.} ~~ & (\mc{A},\mc{H})\mbox{-structural~observability},\\
	~~ &  \sum_{i=1}^{N} \mathcal{H}_{ij} \leq 1, ~~  \sum_{j=1}^{n} \mathcal{H}_{ij} = 1\\
	~~ &  \mc{H}_{ij} \in \{0,1\}\\
	\end{aligned}
\end{equation}

First, notice that $(\mc{A},\mc{H})$ structural observability is equivalent to $(A,H)$ structural observability. Also, note that we added extra conditions; $\sum_{i=1}^{N} \mathcal{H}_{ij} \leq 1$ implies that every social state is observed by at most one agent. $\sum_{j=1}^{n} \mathcal{H}_{ij} = 1$ implies the assumption we made that every agent measures one social state. Therefore, these assumptions do not change the problem and the solution. Next, we apply the fact that parent SCCs are separate components and do not share any state node.  Then, the problem is reformulated as assigning each agents to measure (a state in) a parent SCC.
For this formulation, consider a new cost matrix, denoted by $\bar{c}_{N \times N}$. In this cost matrix we assign the social state with \textit{minimum} cost to be measured by the agent; therefore, $\bar{c}_{ij}$ is the cost of assigning the minimum cost state of  $S^p_j$ to agent $y_i$. Mathematically:
\begin{equation} \label{eqSCCcost}
\bar{c}_{ij}= \min \{c_{il}\},~ x_l \in  S^p_j,~ i,j \in \{1, \hdots, N\}
\end{equation}
In this direction, matrix $c_{N \times n}$ is reformulated as matrix $\bar{c}_{N \times N}$, i.e.  the cost matrix of relating agents to social states is changed to cost matrix of relating agents to parent SCCs. Further, consider a new structured matrix $\mathcal{Z}\sim \{0,1\}^{N \times N}$. This matrix assigns agents to parent SCCs and $\mathcal{Z}_{ij}$ implies agent $i$ is assigned to observe a state in $S^p_j$. Recall that  by sensing (a state in) each parent SCC observability is guaranteed, the problem is reformulated in a new setup as follows:

\begin{equation}
	\begin{aligned}
	\argmin
	\limits_{\mathcal{Z}} ~~  & \sum_{i=1}^{N} \sum_{j=1}^{N} (\bar{c}_{ij}\mathcal{Z}_{ij}) \\
	\text{s.t.} ~~ &  \sum_{j=1}^{N} \mathcal{Z}_{ij} = 1, ~~  \sum_{i=1}^{N} \mathcal{Z}_{ij} = 1 \\
	~~ &  \mathcal{Z}_{ij} \in \{0,1\} \\
	\end{aligned}
	\label{minlsap}
\end{equation}

Note that in above formulation, the constraint $ \sum_{j=1}^{N} \mathcal{Z}_{ij} = 1$ implies that every  parent SCC is observed, and therefore guaranteeing observability. In combinatorial optimization, the formulation \eqref{minlsap}  is referred to as Linear Sum Assignment Problem (LSAP) \cite{assignmentSurvey} and has a  computationally efficient solution which is known as the \textit{Hungarian method} \cite{edmondsHungarian,kuhnHungarian}. The complexity order of this algorithm is $\mathcal{O}(N^3)$. Notice that  formulation~\eqref{minlsap} is a relaxation to original formulation~\eqref{eq_formulation1}, and therefore a polynomial order solution is provided to the sensing cost optimization problem.

It should be mentioned that the assignment reformulation is based on the fact that parent SCCs are disjoint. This is the only requirement on these sets. Therefore, in case that, for a general social digraph, the contractions and SCCs are disjoint the same relaxation and reformulation holds. Therefore if  $|\mc{S}^{p} \cap \mc{C}|= \emptyset$ redefine the modified costs as,
\begin{equation} \label{eqSCCCcost}
\bar{c}_{ij}= \min \{c_{il}\},~ x_l \in  S^p_j ~|~ x_l \in  \mc{C}_j,~ i,j \in \{1, \hdots, N\}
\end{equation}
This cost matrix relates the agents' observations to parent-SCCs or contractions. Then, using a similar relaxation as for parent SCCs, the original  formulation~\eqref{minlsap} is relaxed to assignment formulation~\eqref{eq_formulation1} and Hungarian method is applied.

\subsection{Networking Cost Optimization}
In this subsection, we discuss the problem of designing the structure of  the cyber-network $\mc{G}_W$  such that $(W \otimes A,D_H)$-observability is satisfied. There may be many options of network structures to satisfy distributed observability, where every choice of cyber-network accompanies with a cost.
Assume that all possible links among agents are captured by network $\mc{G}_{net}$, where every link in $\mc{G}_{net}$ has a cost. These costs are referred to as networking cost. The problem can be described as minimizing networking cost, from possible links in $\mc{G}_{net}$, while satisfying distributed observability. Mathematically:
\begin{equation} \label{eq_formulation}
	\begin{aligned}
	\displaystyle
	\argmin
	\limits_{\mc{W}} ~~ &  \sum_{i=1}^{N} \sum_{j=1}^{N} b_{ij}\mc{W}_{ij} \\
	\text{s.t.} ~~ & (W \otimes A,D_H)\mbox{-observability},\\
	~~ & \mc{G}_W \subset \mc{G}_{net},~~   \mc{W}_{ij} \in \{0,1\}\\
	\end{aligned}
\end{equation}
where $\mc{W}$ is the $0-1$ structure of adjacency matrix $W$. In this section we assume that  condition for  observability (based on Theorem~\ref{thm_centralized}) holds. Therefore, \textit{the focus of this section is to satisfy distributed observability constraint by elaborate cost-optimal design of the  cyber-network structure}. This problem is NP-hard to solve, therefore we focus on a special case of this problem; as in Section~\ref{sec_sensingcost} we consider  the social graph to be matched, and the cyber-network $\mc{G}_W$ only contains $\mc{G}_\beta$. Therefore, following the structural results of Theorem~\ref{thm_dist}, the only condition on $\mc{G}_W$ is to be SC \cite{nuno-suff.ness,battistelli2011information,usman_cdc:10}.  Therefore, the original problem \eqref{eq_formulation}, assuming that the centralized  observability conditions are satisfied, is reformulated as,

\begin{equation} \label{eq_formulation2_2}
	\begin{aligned}
	\displaystyle
	\argmin
	\limits_{\mc{W}} ~~ & \sum_{i=1}^{N} \sum_{j=1}^{N} b_{ij}\mc{W}_{ij} \\
	\text{s.t.} ~~ & \mc{G}_W \subset \mc{G}_{net},
	~~  \mc{G}_W~is~SC\\
	~~ &  \mc{W}_{ij} \in \{0,1\}\\
	\end{aligned}
\end{equation}

Here, we discuss the solution for networking cost optimization problem  \eqref{eq_formulation2_2} to find the minimum cost Strongly-Connected (SC) subgraph of network $\mc{G}_{net}$ spanning all nodes (agents). Note that this problem in general represents the Minimum Spanning Strong Subdigraph and is NP-hard \cite{digraphs}. In \cite{frederickson1981approximation}, a 2-approximations heuristic algorithm to solve this problem is proposed. However, \textit{under the assumption that the links in the cyber-network are bidirectional this problem has a polynomial order solution.} In other words, agents share their information mutually and $W$ matrix is symmetric.
Then the networking cost optimization problem is relaxed to a known problem in combinatorial optimization known as the  Minimum  Spanning Tree (MST) or Minimum Weight Spanning Tree  \cite{gabow1986efficient}. It is known that this problem has a polynomial order solution using \textit{Prim's algorithm} \cite{prim1957shortest} or \textit{Kruskal's algorithm} \cite{kruskal1956shortest} with computational complexity of $\mc{O}(N^2)$.

\section{Simulations and Illustrative Examples} \label{sec_sim}
\subsection{Simulation of Distributed Inference}\label{sec_sim1}
\begin{figure}
	\centering
	{\includegraphics[width=3.3in]{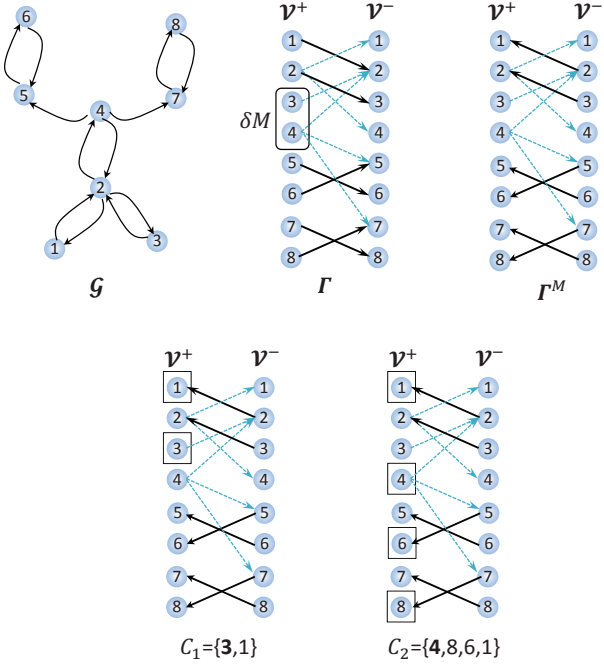}}
	\caption{ This figure shows a social digraph,~$\mc{G}$ and its bipartite representation,~$\Gamma$. In $\Gamma$ the black edges are matching edges, and $\delta \mc{M}$ represents the unmatched nodes. By reversing the matching edges, the auxiliary graph,~$\Gamma^{\mc{M}}$ is constructed. In~$\Gamma^{\mc{M}}$  following the alternating paths the contractions $\mc{C}_1$ and $\mc{C}_2$ are found.}
	\label{fig_graph}
\end{figure}

Consider a social graph, $\mc{G}$, of $8$ social states as shown in Fig.~\ref{fig_graph}. Note that this social digraph is structurally rank-deficient. We have $S\mbox{-rank}(\mc{A})=6$.
To find the contractions in digraph, following Section~\ref{sec_cont}, define its associated bipartite graph~$\Gamma$, matching $\mc{M}$, unmatched nodes $\delta \mc{M}$, auxiliary graph $\Gamma^\mc{M}$, and finally contractions are shown in Fig.~\ref{fig_graph}.

\begin{figure}
	\centering
	{\includegraphics[width=3in]{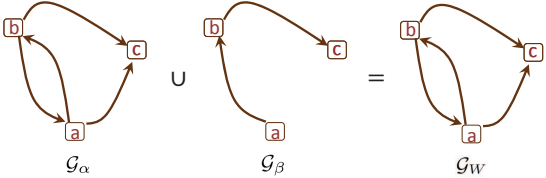}}
	\caption{ This figure show the structure of $\mc{G}_\alpha$ over which agents share their measurements and $\mc{G}_\beta$ over which agents share their state predictions. The combination of these two makes the cyber-network $\mc{G}_W$. }
	\label{fig_W}
\end{figure}

Contractions $\mc{C}_1=\{x_3,x_1\}$ and $\mc{C}_2=\{x_4,x_8,x_6,x_1\}$ are shown in the figure. There are two parent SCCs in this social graph, $\mc{S}^p_1=\{x_5,x_6\}$ and $\mc{S}^p_1=\{x_7,x_8\}$\footnote{Note that in this example parent SCCs are only consist of one state node, and therefore the SCCs are indeed self-cycles.}.
For observability measurement of a state in all parent SCCs and contractions are sufficient. We choose to measure states $x_3$ and $x_6$ by (Type-$\alpha$) agents $a$ and $b$, and measure state $x_7$ by (Type-$\beta$) agent $c$. For minimum number of agents (see Section~\ref{sec_minobsrv}) we observe state $x_6$ to recover both accessibility of $\mc{S}^p_1$ and S-rank of $\mc{C}_2$. Next, for (distributed) inference, we design the structure of cyber-network $\mc{G}_W$. This is based on Theorem~\ref{thm_dist}. As shown in Fig.\ref{fig_W} this network is a combination of $\mc{G}_\alpha$ and $\mc{G}_\beta$.

\begin{figure}
	\centering
	{\includegraphics[width=2.5in]{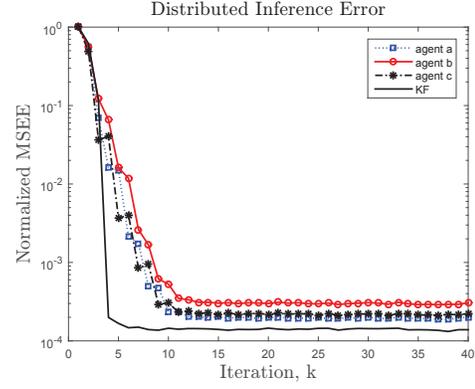}}
	\caption{This figure shows the inference error evolution in time for social graph of Fig.\ref{fig_graph} monitored by the agents connected through the cyber-network of Fig.\ref{fig_W}. The protocol follows the equations \eqref{eq_p}-\eqref{eq_m} and the error dynamics follows equation \eqref{eq_err1}. As it can be seen from the figure, the error at each agent is bounded steady-state. For comparison, the MSEE for estimation based on \textit{centralized} Kalman Filter (KF) is given by black line. As expected the centralized estimation gives lower MSEE because all measurement information and global observability are available at the estimator.}
	\label{fig_simulation}
\end{figure}

Agents share their measurements over  $\mc{G}_\alpha$ and share their predictions over $\mc{G}_\beta$. For simulation the social graph weights in social dynamic equation~\eqref{eq_sys1} are considered randomly such that $\rho (A)=1.2>1$. This is to avoid trivial solution for inference. The weights for $W$ matrix associated with $\mc{G}_\beta$ are considered random such that $W$ is stochastic. We intentionally choose the entries of $A$ and $W$ randomly, to show that the results are generic and irrespective of the particular weights of fusion rule among agents and the social dynamic parameters.
Using the inference protocol~\eqref{eq_p} and~\eqref{eq_m}, the inference error follows the dynamic equation~\eqref{eq_err1}. Noise terms for system and measurement are considered as
$\mc{N}(0,0.01)$. The evolution of agents' Mean Squared Estimation Error (MSEE) in distributed inference, averaged over $1000$ Monte-Carlo simulations, are shown in Fig.~\ref{fig_simulation}.
The design of block-diagonal gain matrix $K_k$ in equation~\eqref{eq_err1} is based on Linear Matrix Inequality (LMI) approach. The  gain matrix being block-diagonal supports the fact that the proposed inference methodology is distributed. The steady-state bounded MSEE verifies our distributed inference.  \textit{Centralized} Kalman Filter (KF) \cite{kalman:61} is also provided.

As compared to the works on the distributed inference in the literature, the number of  links in the cyber-network $\mc{G}_W$ (Fig. \ref{fig_W}) is less than the inference method in \cite{sauter:09,commault_recovery,kar2012distributed}. In \cite{sauter:09,commault_recovery}, there is no classification of Type-$\beta$ and $\alpha$ agents and both types are required to share their information directly with others. Therefore, the cyber-network must be all-to-all network which in general requires more links in the cyber-network as compared to our work. Similarly, \cite{kar2012distributed} requires global observability in the neighborhood of each agent and therefore more connectivity in the cyber-network. This difference in number of links is more significant for large-scale systems. For semi-centralized case, Kalman Estimator may be applied which gives the best performance in terms of MSEE. This is because all the measurements and global observability are  available at the estimator, while in distributed case the observability is local and partial measurements are available at the neighborhood of each agent.
Further, other distributed inference literature 	\cite{nuno-suff.ness,battistelli2011information,usman_cdc:10,Sayed-LMS} are restricted to the full-rank systems and therefore are useless in scenarios where the social digraph is rank-deficient as in Fig.  \ref{fig_graph}, and therefore, we are unable to make any comparison on the MSEE performance with these works. This difference  makes the contribution of our work significant as compared to the similar distributed inference literature \cite{nuno-suff.ness,battistelli2011information,usman_cdc:10,Sayed-LMS}.
	
\subsection{Illustration of Cost-Optimal Design}
\begin{figure}[!t]
	\centering
	\includegraphics[width=2.5in]{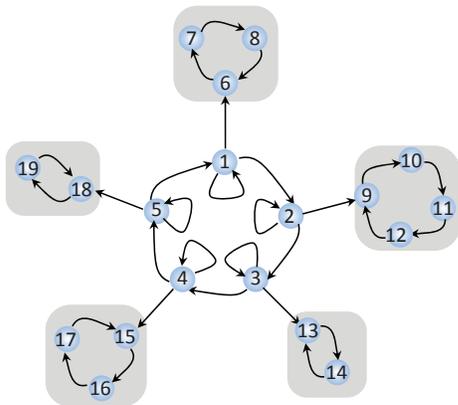}
	\caption{This figure shows a social digraph of $19$ states. It contains $6$ SCCs, determined by DFS algorithm. The highlighted components have no incoming edge from other components and therefore are parent SCCs.}
	\label{figgraph2}
\end{figure}

\begin{figure}[!t]
	\centering
	\includegraphics[width=2.5in]{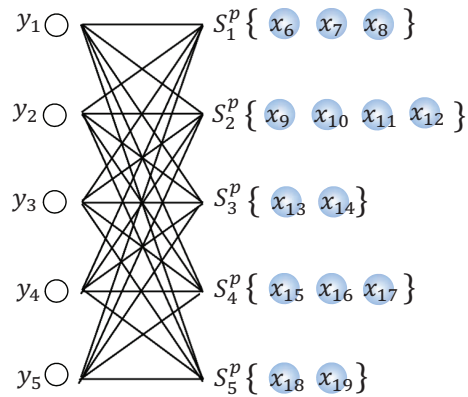}
	\caption{This figure shows the assignment of $5$ agents to measure social states in  $5$ parent SCCs (for Fig.\ref{figgraph2}). The assignment cost for each parent SCC is according to \eqref{eqSCCcost}. The implication is that each agent is assigned with the minimum cost of measuring social states in each SCC. This frames  the problem as a LSAP (equation \eqref{minlsap}) and solvable by Hungarian method.}
	\label{figassign2}
\end{figure}

In this subsection, we provide an example to illustrate the results of Section~\ref{sec_optimal} on LSAP and MST formulation. We consider a social graph example similar to the example provided in \cite{pequito_gsip}. This example includes a social graph of $19$ states as shown in Fig.~\ref{figgraph2}.
Now, consider a group of agents monitoring this social graph. We assume random costs for agents measuring different states. Note that this social graph is structurally matched. Using the DFS algorithm $5$ parent SCCs are found in this social graph in $\mc{O}(n^2)$ as shown in Fig.\ref{figgraph2}. Therefore, according to Section~\ref{sec_distributed}, for observability and inference it is sufficient that one state in each parent SCC be observed by an agent.   Assume that each state observation is assigned with a cost $c_{ij}$. Each agent chooses the minimum cost state in the assigned parent SCC. Then, the problem  is  to optimally assign agents to parent SCCs as shown in Fig.~\ref{figassign2}. This problem is indeed in the form of \eqref{minlsap} and can be solved by the Hungarian method in $\mc{O}(N^3)$. We intentionally consider the social network example in Fig. \ref{figgraph2} as it is similar to the example given in \cite{pequito_gsip}. In \cite{pequito_gsip} it is claimed that for such social network the sensing cost optimization problem is NP-hard to solve. However, in this section we provide a polynomial-order solution of $\mc{O}(n^2+N^3)$ to optimize the sensing cost of the social digraph. This is significant as polynomial-order solutions are applicable in large-scale cyber-social networks.

Next we consider networking cost optimization over the cyber-network of agents measuring states in $\{\mc{S}^p_1,...,\mc{S}^p_5\}$. We consider $\mc{G}_{net}$ to be all-to-all network of $5$ agent nodes and its cost matrix $b$ (as link weights) to be a symmetric random matrix as follows:
\begin{eqnarray}\nonumber \small
\left(
\begin{array}{ccccc}
1.5155  &  1.4492  &  1.2157  &  1.0942  &  0.6610\\
 1.4492  &  0.0637 &   0.9718 &   0.4277   & 0.5427\\
1.2157  &  0.9718 &   0.6342  &  1.7157  &  0.6808\\
1.0942  &  0.4277  &  1.7157  &  1.5904  &  0.8962\\
0.6610  &  0.5427  &  0.6808  &  0.8962  &  1.5094
\end{array}
\right)
\end{eqnarray}

\begin{figure}[!t]
	\centering
	\includegraphics[width=1.8in]{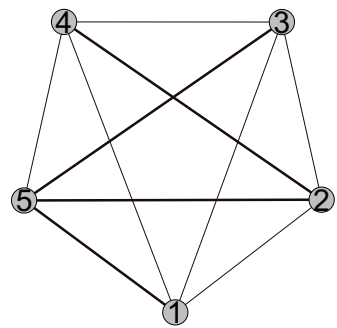}
	\caption{This figure shows the all-to-all network $\mc{G}_{net}$ representing all possible communications among agents in the cyber-network. Each communication link has a cost. The networking cost optimization  finds $\mc{G}_W$ as the SC subnetwork with minimum cost, which is known to be the minimum cost spanning tree.  The bold links represent such $\mc{G}_W$. }
	\label{fig_network}
\end{figure}

Following the results of Section~\ref{sec_distributed} for distributed observability the network connecting these agents is sufficient to be SC. Among the possible solutions to find SC $\mc{G}_W$, the equation~\eqref{eq_formulation2_2} and  MST algorithm finds the one with minimum weight as the solution based on Prim's algorithm  or Kruskal's algorithm in $\mc{O}(N^2)$. The solution for this example is shown in Fig.~\ref{fig_network}. We remind the reader that the networking cost optimization problem generalizes the centralized cost-optimal inference problem in literature. In \cite{pequito2017structurally} the cost-optimal design of the communications to the central base is considered, which is only applicable in centralized inference scenarios. To the best of our knowledge, there are no cost-optimal networking designs based on distributed inference in the literature for the sake of comparison. In this section, we provide a polynomial order method of $\mc{O}(N^2)$ to solve the networking cost optimization problem, which is applicable in distributed inference of large-scale cyber-social systems.

\section{Concluding Remarks}\label{sec_concluding}
This paper formally develops the modeling, distributed inference, and cost-optimal design of the cyber-social systems.
We adopt a structural (or generic) approach independent of the particular dynamics over the social network and particular fusion rule,  e.g., Metropolis-Hastings~\cite{Xiao05distributedaverage}, as inference protocol. This is better clarified in simulations, as we choose the entries of system matrix and cyber-network to be \textit{random}. Therefore, the inference protocol works for any choice of system dynamics and inference fusion rule. As compared to the distributed inference protocols in the literature which are only applicable when the underlying dynamic system is full-rank, our protocol is independent of the rank of the system. The distributed inference over rank deficient example in Section~\ref{sec_sim1} verifies this novelty of our work.  Next, we develop minimum observability requirement of the social digraph and further design the structure of cyber-network to minimize the sensing and networking cost for distributed inference. This is significant as it generalizes the centralized networking cost optimization in the literature and also finds a polynomial order solution for claimed NP-hard sensing cost optimization.
The optimization solution and other algorithms in this paper are of polynomial order complexity. To summarize, the contraction detection algorithm and the DFS algorithm are, respectively, $\mc{O}(n^{2.5})$ and $\mc{O}(n^2)$, where $n$ is the size of the social system. The LSAP solution (the Hungarian method) and the MST solution (the Prim's algorithm) are, respectively, $\mc{O}(N^3)$ and $\mc{O}(N^2)$, where $N$ is the size of the cyber-network. Low computational complexity of these algorithms is one reason to adopt LSI model over numerical counterparts, e.g. LTI model. 

Other points to mention here is on volatile social networks with changing links. In case of link addition in the social network the number of agents and their connectivity in the cyber-network  is still sufficient for inference, however link removal may dictate more agents and therefore more connectivity in the cyber-network. This is the topic of our future research as one may extend the results of this paper to time-varying structures of social systems. Further, in terms of network sparsity, lower rank sparse networks  have more unmatched nodes and therefore require more Type-$\alpha$ agents. This, in turn, imposes more connectivity  on the cyber-network of agents. It should be noted that, in general, the load of communications and data processing among agents is not balanced and depends on the position of the agent in the cyber-network. In general, Type-$\alpha$ agents are accompanied with more communication loads.  As another direction of future research, one may consider noisy channels and packet loss in the communication links, link failure, and even cyber-attacks on the communication links in the cyber-network.

\bibliographystyle{IEEEbib}
\bibliography{bibliography}

\end{document}